\documentclass{article}

\usepackage{arxiv}

\usepackage[utf8]{inputenc} 
\usepackage[T1]{fontenc}    
\usepackage{hyperref}       
\usepackage{url}            
\usepackage{booktabs}       
\usepackage{amsfonts}       
\usepackage{nicefrac}       
\usepackage{microtype}      
\usepackage{lipsum}
\usepackage{natbib}

\usepackage{amsmath}
\usepackage{multirow}
\usepackage{booktabs}
\usepackage{hyperref}
\usepackage{graphicx}

\newcommand{\bs}{\boldsymbol}
\newcommand{\bD}{\bs{D}}
\newcommand{\bI}{\bs{I}}

\newcommand{\bR}{\bs{R}}
\newcommand{\bX}{\bs{X}}

\newcommand{\bU}{\bs{U}}
\newcommand{\bV}{\bs{V}}
\newcommand{\bY}{\bs{Y}}

\newcommand{\bbeta}{\bs{\beta}}
\newcommand{\sigsq}{\sigma ^ 2}
\newcommand{\tausq}{\tau ^ 2}
\newcommand{\bSigma}{\bs{\Sigma}}
\newcommand{\bmu}{\bs{\mu}}
\newcommand{\btheta}{\bs{\theta}}
\newcommand{\bTheta}{\bs{\Theta}}
\newcommand{\bOmega}{\bs{\Omega}}

\newcommand{\mcal}{\mathcal}

\newcommand{\mD}{\mcal{D}}
\newcommand{\mM}{\mcal{M}}
\newcommand{\mT}{\mcal{T}}

\newcommand{\Ystar}{Y^*}

\newcommand{\supb}{^{(b)}}
\newcommand{\bepsilon}{\bs{\epsilon}}

\title{Borrowing from Supplemental Sources to Estimate Causal Effects from a Primary Data Source}

\author{
  Jeffrey A. Boatman \\
  Division of Biostatistics\\
  University of Minnesota\\
  Minneapolis, MN 55455 \\
  \texttt{boat0036@umn.edu} \\
   \And
 David M. Vock \\
  Division of Biostatistics\\
  University of Minnesota\\
  Minneapolis, MN 55455 \\
  \texttt{vock@umn.edu} \\
  \And
  Joseph S. Koopmeiners \\
  Division of Biostatistics\\
  University of Minnesota\\
  Minneapolis, MN 55455 \\
  \texttt{koopm007@umn.edu} \\
}

\begin{document}
\maketitle

\begin{abstract}
The increasing multiplicity of data sources offers exciting possibilities in estimating the effects of a treatment, intervention, or exposure, particularly if observational and experimental sources could be used simultaneously. Borrowing between sources can potentially result in more efficient estimators, but it must be done in a principled manner to mitigate increased bias and Type I error. Furthermore, when the effect of treatment is confounded, as in observational sources or in clinical trials with noncompliance, causal effect estimators are needed to simultaneously adjust for confounding and to estimate effects across data sources. We consider the problem of estimating causal effects from a primary source and borrowing from any number of supplemental sources. We propose using regression-based estimators that borrow based on assuming exchangeability of the regression coefficients and parameters between data sources. Borrowing is accomplished with multisource exchageability models and Bayesian model averaging. We show via simulation that a Bayesian linear model and Bayesian additive regression trees both have desirable properties and borrow under appropriate circumstances. We apply the estimators to recently completed trials of very low nicotine content cigarettes investigating their impact on smoking behavior.  

\end{abstract}

\keywords{Bayesian Additive Regression Trees \and Bayesian Linear Model \and Bayesian Model Averaging \and Borrowing \and Causal Inference \and Multisource Exchangeability Models.
}

\section{Introduction} \label{introduction}

  The Center for Evaluation of Nicotine in Cigarettes (CENIC) conducts randomized controlled trials (RCTs) investigating how marked decreases in the nicotine content of cigarettes affect the use and effects of tobacco in current smokers. Their work is part of a body of research meant to inform the United States Food and Drug Administration (FDA) as it contemplates strategies to reduce the public health burden of smoking. Not surprisingly, because smokers enrolled in trials of very low nicotine content (VLNC) cigarettes  still have access to normal commercial cigarettes, noncompliance to randomized treatment in these trials is very common \citep{Nardone2016}. The intention to treat (ITT) analysis reported for these trials thus estimates the effect of nicotine reduction if smokers still have access to commercial cigarettes. But ITT does not estimate the effect of a potential FDA-mandated reduction in the nicotine content of cigarettes: in this hypothetical future, normal nicotine cigarettes would not be legally available, and all smokers would be forced to comply with a nicotine reduction. Causal effect estimators addressing noncompliance must therefore be used in conjunction with ITT estimators to fully understand the possible effects of FDA regulations on the nicotine content of cigarettes. Unfortunately, estimating causal effects can be inefficient if noncompliance is high or in the presence of strong confounding. 
  
  One way to improve efficiency is to combine data sources for causal inference. 
   This is a particularly attractive option for CENIC given  similarities in entrance criteria and interventions across studies. CENIC project 1, study 2 (CENIC-P1S2)
   was a 6-week, randomized, $2\times2$ factorial design to evaluate the effect of VLNC versus normal nicotine content (NNC) cigarettes in the presence or absence of transdermal nicotine (TDN)\citep{Smith2019}.
  Each treatment group is small ($n \approx 60$/group), posing further challenges for estimating causal effects, but extensive supplementary data are available from CENIC project 1, study 1 (CENIC-P1S1) \citep{Donny2015} and from CENIC project 2 (CENIC-P2)\citep{Hatsukami2018}. CENIC-P1 was a 6-week RCT that randomized smokers to one of seven groups consisting of usual brand cigarettes or experimental cigarettes with nicotine content ranging from NNC to VLNC cigarettes (0.4 mg nicotine per gram tobacco). From this trial, data to supplement CENIC-P1S2 include the usual brand cigarette group as control and the VLNC groups as treatment (high and low tar groups combined). CENIC-P2 was a 20-week RCT that randomized smokers to one of three groups, 1) maintenence on NNC cigarettes, 2) immediate reduction to VLNC cigarettes, of 3) gradual reduction to VLNC cigarettes. From this trial, data to supplement CENIC-P1S2 include the NNC group as control and the immediate-reduction group as treatment using outcomes collected at week 8. 

 
  One approach to combining data is simply to pool data from multiple sources to increase the overall sample size. This naive approach would increase the precision of the estimated effect of an intervention, but it leads to both bias and inflated type 1 error rates in the presence of inter-source heterogeneity. A more statistically rigorous approach is to use estimators that ``shrink" the estimated treatment effect from a primary data source towards estimates of the treatment effect from supplementary sources, thus borrowing strength and increasing precision. 
  For example, we will have more confidence in our estimate of the treatment effect in a small trial if it is consistent with the treatment effect observed in the supplemental data source. 

  Statistical methods that facilitate the borrowing of information across data sources arise most naturally in the Bayesian paradigm. These methods have been used in the meta-analysis literature and have more recently been adapted to the analysis of RCTs. Bayesian approaches to incorporating supplemental information into the analysis of RCTs include the derivation of predictive prior distributions using the standard hierarchical model
  \citep{Gelman2006}; power priors, which use a pre-specified parameter that down-weights supplemental data \citep{Ibrahim2000}; and commensurate priors, which utilize a hierarchical modeling framework with a commensurability parameter to control borrowing from supplemental sources
  \citep{Hobbs2011}.

  \citet{Kaizer2018} recently introduced multi-source exchangeability models (MEMs), a novel approach to integrating supplementary data into the analysis of a primary data source. 
  MEMs offer a flexible approach to cross-source data integration by including parameters that separately control borrowing from each supplementary data source and can be estimated from the data. This can be achieved by specifying models that represent all possible ways that the supplementary data can be combined with the primary data source and using Bayesian model averaging to obtain a weighted average of the estimated treatment effect across all models. Simulation results suggest that MEMs have favorable statistical properties when compared to existing hierarchical modeling strategies \citep{Kaizer2018}.

  In this manuscript, we show how to use MEMs to borrow from supplemental sources in the estimation of causal effects using a primary data source. Our method uses regression models to estimate causal effects in the presence of confounding. 
  In Section \ref{methods}, we discuss two models for estimating causal effects, a Bayesian Linear Model (BLM), and Bayesian Additive Regression Trees (BART)\citep{Chipman2010}, and we should how to use these models to borrow from supplemental data to estimate causal effects from a primary data source in the MEMs framework. In Section \ref{simulation}, we present results of a simulation study to assess the performance of the estimators. In Section \ref{application} we estimate the causal effect of VLNC cigarettes from CENIC-P1S2 allowing borrowing from CENIC-P1S1 and CENIC-P2, and we conclude in Section \ref{discussion}.

\section{Methods} \label{methods}

\subsection{Observed Data}

We consider data from a single primary source $P$ and $H$ supplemental sources with sample sizes $n_P, n_1, \dots, n_H$. The data are assumed to be random samples from each source population. We assume the following data are collected for the $i$th participant: a source variable $S_i$ that can take values $P, 1, \dots, H$, indicating the data source; the quantitative outcome $Y_{i}$; a binary indicator of treatment assignment $A_{i}$, with untreated (or control) participants having $A_{i} = 0$ and $A_{i} = 1$ indicating assignment to treatment; and a vector of patient characteristics $\bX_{i} = \left(X_{i1}, \dots, X_{ip}\right) ^ T$ associated with $Y_{i}$ and possibly with $A_{i}$ if treatment is not randomly assigned. 
For data source $s$, let $\mD_s = \left\{\left(Y_i, A_i, \bX_i, S_i \right): S_i = s\right\}$ represent the observed data. $\mD = \left(\mD_P, \mD_1, \dots, \mD_H\right)$ denotes the collection of data from all sources.

\subsection{Potential Outcomes and Target of Inference}
Let $\Ystar(a, s)$ denote the outcome of a randomly selected participant if, possibly contrary to fact, $A = a$ and $S = s$. Because for each participant we do not observe $\Ystar(a, s)$ for all $a$ and $s$, $\Ystar(a, s)$ is known as a potential outcome. Our target of inference is the population average treatment effect (PATE) $ \Delta_P = E_{\bX | S = P}\left[E \left\{\Ystar(1, P) - \Ystar(0, P) | \bX, S = P\right\}\right]$, the expected difference difference for being treated versus remaining untreated in the primary source population, where the outer expectation is with respect to the conditional distribution of $\bX|S = P$.


\subsection{Identifying Assumptions} \label{sec:assumptions}
  
  We make the following assumptions to relate the distribution of the observed data to the distribution of the potential outcome \citep{robins2008}. First, we assume that the potential outcome under treatment $a$ in source $s$ is equal to the observed outcome given that $A = a$ and $S = s$, which can be restated as $Y = \sum_{s = P, 1, \dots, H} \left\{A \Ystar\left(1, s\right) + \left(1 - A\right) \Ystar\left(0, s\right)\right\}I\left(S = s\right)$. This is referred to as the consistency assumption. Second, we assume that we have collected sufficient data on participant characteristics so that treatment assignment is conditionally ignorable, or $\left.\left\{A \perp \Ystar(0, S), \Ystar(1, S)\right\} \;\middle\vert\; \bX, S\right.$. Third, we assume that $0 < Pr \left(A = 1 | \bX,  S\right)< 1$ for all $\bX$. This is referred to as the positivity assumption.
  
\subsection{Estimators}
  
  Under the assumptions stated above, we have the following equalities: $E \left(Y | A = 1, S, \bX\right) = E \left\{\Ystar(1, S) | S, \bX\right\}$, and $E \left(Y | A = 0, S, \bX\right) = E \left\{\Ystar(0, S) | S, \bX\right\}$. This allows us to work with the observed data only and to estimate $\Delta_P$, provided that we have an estimator for $E \left(Y | A, S, \bX\right)$.
  Although many estimators are possible, we describe two, a Bayesian linear model (BLM) and Bayesian additive regression trees (BART). For now, we consider the simple case where the only data available are from the primary source, and there is no borrowing.  We discuss borrowing from supplemental sources in Section \ref{sec:mems}. 
  

\subsubsection{Bayesian Linear Model}

  The first estimator we consider is the Bayesian linear model (BLM) with a normal-inverse-gamma prior. Define the model matrix $\bD$ with the $i$th row equal to $\left(1, A_i, \bX_i^T\right) ^ T$, and let $\bY = \left(Y_i, \dots Y_{n_P}\right) ^ T$ denote the outcome vector. The BLM is
  \begin{equation}
    \bY = \bD \bbeta_P + \bepsilon, \quad \bepsilon \sim N\left(0, \sigsq_P\bI\right),    
    \label{eqn:bayes_lm}
  \end{equation}
  with 
  $
    \sigsq_P \sim IG\left(a_P, b_P\right),
    \bbeta_P | \sigsq_P \sim N\left(\bmu_P, \sigsq_P \bV_P\right)
  $, where $a_P, b_P, \bmu_P, \text{ and } \bV_P$ are known hyperparameters. The subscript $P$ on the coefficients and parameters is meant to clearly indicate that these are for modelling data in the primary source $P$.
  
  Our hyperparameter specification is similar to those of \citet{Raftery1997}. We let $\bmu_P = \left(\Bar{y}, 0, \dots, 0\right) ^ T$ where $\Bar{y}$ is the sample mean of $Y_1, \dots, Y_{n_P}$, and $\bV_P = \left(\frac{1}{n_P}\bD ^ T \bD\right)^{-1}$. We choose $a_P$ and $b_P$ so that $\sigsq_P$ has mean $\hat{\sigma}^2$ and variance $2\hat{\sigma}^4$, where $\hat{\sigma}^2$ is from the least squares regression of $\bY$ on $\bD$. 
  
  The primary advantage of the BLM is the resulting efficiency of the estimator for $\Delta_P$. If Model (\ref{eqn:bayes_lm}) were the correct model, and if $\bbeta$ were estimated using the ordinary least squares (OLS) estimator $\hat{\bbeta}$, then the estimator for $\Delta_P$ implied by $\hat{\bbeta}$ would be consistent and the most efficient \citep{lunceford2004}. Although the efficiency of the Bayesian estimator for $\Delta_P$ using Model (\ref{eqn:bayes_lm}) depends on the prior specification, the estimator will have performance asymptotically equivalent to the OLS estimator provided that a suitable posterior summary statistic (e.g., the mean) is chosen as the estimator for $\Delta_P$. The primary disadvantage of using of Model  (\ref{eqn:bayes_lm}) is that specifying a correct, or approximately correct, model is challenging if the dimension of $\bX$ is large, particularly if interactions or nonlinear relationships are present. 
 
  \subsubsection{Bayesian Additive  Regression Trees}
  \label{sec:BART}
    BART \citep{Chipman2010} is a nonparametric tree ensemble model primarily used for prediction. Prior to fitting the model, \citet{Chipman2010} shift and scale $Y$ so that the sample mean is 0 and the range is 1; we assume throughout this section that $Y$ is on the transformed scale. BART consists of two components, a sum-of-trees model, and a regularization prior. The sum-of-trees model with $m$ trees is 
    \begin{equation}
      Y_i = \sum_{j=1}^m g\left(A_i, \bX_i; T_{Pj}, M_{Pj}\right) + \epsilon_i, \quad \epsilon_i \sim N\left(0, \sigsq_P\right).
      \label{eqn:BART-model}
      \end{equation}
  where $T_{Pj}$ denotes the $j$th tree and $M_{Pj} = \left(\mu_{Pj1}, \dots, \mu_{Pjk_{j}}\right)$ denotes the terminal node values of $T_{P_j}$.
  The function $g$ 
  sends $A$ and $\bX$ down branches of each of the $m$ trees and sorts it based on interior node binary decision rules of the form $x \le c$ vs. $x > c$ until a terminal node is reached. Each terminal node then assigns each observation one value of  $M_{Pj}$.  
  Thus the conditional mean value of $Y$ is the sum of the terminal node values to which the observation has been assigned based on values of $A$ and $\bX$. The regularization prior ensures that each tree is a ``weak learner'' and contributes a small amount to the overall fit. The terminal node values are {\it a priori} independent and identically distributed as $N \left(0, \tausq\right)$, where $\tausq$ is a constant. The prior for $\sigsq_P$ is an inverse gamma, $\sigsq_P \sim  IG\left(\frac{\nu}{2}, \frac{\nu\lambda}{2}\right)$, with $\nu$ and $\lambda$ both fixed. \citet{Chipman2010} use a data-informed approach to select $\lambda$, and they recommend a default value of $\nu = 3$. See \citet{Chipman2010} for further details on the regularization prior and specification of hyperparameters. 
  
  \citet{Hill2011} proposed using BART for causal inference due its ease of use, its flexibility in modelling complex response surfaces, and its ability to identify heterogeneous treatment effects. Similar to the BLM, the analyst must specify the outcome, treatment, and confounders. But unlike the BLM, the analyst need not specify a functional form for the relationship between the outcome and the predictors. Any interactions and non-linear effects that could escape the attention of the analyst using a linear model can be captured quite naturally through the fitted sum-of-trees model. 
  The primary disadvantage of BART is that it cannot capitalize on linear relationships to accurately predict the potential outcome response surface where very little data are available. BART is therefore not expected to perform well in scenarios with strong confounding.  The posterior credible intervals are naturally wider in these areas, however, which partially ameliorates this problem \citep{Hill2011}. 
  
 \subsection{Posterior Inference}
 \label{sec:posterior-inference}

  Here we describe  posterior inference on $\Delta_P$ assuming we have already fit a model to the primary data source using either BART or the BLM for $E\left(Y|A, S = P, \bX, \btheta_P\right)$ with model parameters $\btheta_P$. For the BLM, $\btheta_P = \left(\bbeta_P, \sigsq_P\right)$; for BART, $\btheta_P = \left\{\left(T_{Pj}, M_{Pj}\right)_{j = 1, \dots, m}, \sigsq_P\right\}$. Suppose we take $B$ draws from the posterior distribution of $\btheta_P$, with  $\btheta_P\supb$ denoting the $b$th draw. For each draw from the posterior, we compute for each $\bX$ in the primary data source the conditional average treatment effect,
  \begin{equation}
    E\left\{Y | A = 1, S = P, \bX, \btheta_P\supb\right\} - 
    E\left\{Y | A = 0, S = P, \bX, \btheta_P\supb\right\},
    \label{eqn:cate}
  \end{equation}
  which is the difference in expected potential outcomes if the observation were assigned treatment versus control. 
  The PATE for the $b$th posterior draw is then 
  \begin{equation*}
      \Delta_P\supb = E_{\bX|S=P}\left[E\left\{Y | A = 1, S = P, \bX, \btheta_P\supb\right\}-E\left\{Y | A = 1, S = P, \bX, \btheta_P\supb\right\}\right].
  \end{equation*}
  The outer expectation requires an integration over $\bX|S=P$, the distribution of covariates in the primary source population. To avoid specifying a multivariate distribution for $\bX|S=P$ with priors, we approximate the integral using the Bayesian bootstrap \citep{Rubin1981}, which has previously been used to integrate over the distribution of covariates \citep{Wang2015, Nethery2019}. For each value of the conditional average treatment effects in (\ref{eqn:cate}), we perform a single bootstrap iteration as follows. Let $\bs{p}\supb = \left\{p_1\supb, \dots, p_{n_P}\supb  \right\}$ denote the vector of sampling probabilities for the $b$th posterior draw as described by \citet{Rubin1981}.  We let $\Delta_P\supb = \sum_{i = 1}^{n_P} p_i\supb \left[E\left\{Y | A = 1, S = P, \bX, \btheta_P\supb\right\} - 
    E\left\{Y | A = 0, S = P, \bX, \btheta_P\supb\right\}\right]$. The resulting collection  $\{\Delta_P^{(1)}, \dots, \Delta_P^{(B)} \}$ is taken to be the posterior distribution of $\Delta_P$.

\subsection{Multisource Exchangeability Models}
\label{sec:mems}

  Having described estimators for the $\Delta_P$ using the primary source only, our goal is now to incorporate information from supplemental data sources into making inference about $\Delta_P$. To do this, we use Multisource Exchangeability Models (MEMs), a Bayesian model averaging (BMA) technique first proposed by  \citet{Kaizer2018}. We begin by selecting a model, either BLM or BART. Let $\btheta_h$ denote the source-specific model parameters for the supplemental data source $h$. We can borrow information from supplemental data source $h$ by assuming that the data source is exchangeable with the primary data source. In this context, data source $h$ is exchangeable with $P$ if the model parameters are identical to the model parameters for the primary data source $P$. As each of the $H$ data sources can be exchangeable, or not, with the primary data source, there are $2 ^ H$ possible patterns of exchangeability. Let $Z_h$ denote an indicator of whether source $h$ is exchangeable with source $P$. The $q$th pattern of exchangeability, corresponding to the $q$th MEM, is denoted as $\bOmega_q = \left(Z_1 = z_{1q}, \dots, Z_H = z_{Hq}\right)$. The model parameters for source $h$ under $\bOmega_q$ are $z_{1q}\btheta_P + \left(1 - z_{1q}\right)\btheta_h$. For simplicity, the priors for $\btheta_P, \btheta_1, \dots, \btheta_H$ are independent so that the prior for $\bTheta_q$, the parameters of $\bOmega_q$, is $p\left(\bTheta_p|\Omega_p\right) = p\left(\btheta_P\right)\prod_{h = 1} ^ H  p\left(\btheta_h\right)^{\left(1 - z_{hp}\right)}$.

  The marginal likelihood is obtained by averaging the likelihood over the prior: 
  \begin{equation}
    p\left(\mD | \bOmega_q\right) = 
    \int 
    p\left(\mD | \bTheta_q, \bOmega_q\right)
    p\left(\bTheta_q | \bOmega_q\right) 
    d\bTheta_q.
    \label{eqn:marginal-likelihood}
  \end{equation}
  For simplicity, we do not specify a model for the joint distribution for $Y, A, S, \bX$, but only for the conditional distribution of $Y$. Thus the likelihood $p\left(\mD|\bTheta,\bOmega_p\right)$ in Equation (\ref{eqn:marginal-likelihood}) is simply a product of conditional normal densities given the model for $Y$. 
  
  The posterior probability that $\bOmega_q$ is the correct model, given the data, is 
  \begin{equation*}
      \omega_q = p\left(\bOmega_q | \mD\right) = 
      \frac{p\left(\mD | \bOmega_q\right)p\left(\bOmega_q\right)}{\sum_{w = 1}^{2 ^ H}p\left(\mD | \bOmega_w\right)p\left(\bOmega_w\right)},
  \end{equation*}
  where $p\left(\bOmega_p\right)$ is the prior probability that $\bOmega_p$ is the true model. We explore prior probabilities on the MEMs in simulation studies of Section \ref{simulation}. 
  
  Finally, for each MEM, data sources that share $\btheta_P$ are simply combined for estimation as they would be in a traditional analysis. Borrowing is thus accomplished, under each MEM, by combining the exchangeable data sources to derive the MEM-specific model parameter posterior distributions for $\btheta_P^{\bOmega_1}, \dots, \btheta_P^{\bOmega_{2^H}}$ and the  concomitant MEM-specific PATE posteriors for $\Delta_P^{\bOmega_1}, \dots, \Delta_P^{\bOmega_{2^H}}$.The posterior for $\Delta_P$ is then a weighted average of the posterior distributions given each MEM, 
  \begin{equation}
      p\left(\Delta_P | \mD\right) = \sum_{q = 1} ^ {2 ^ H} \omega_q p\left(\Delta_P^{\bOmega_q} | \mD, \bOmega_q\right).
      \label{eqn:weighted-posterior}
  \end{equation}
  Draws from the posterior of $p\left(\Delta_P^{\bOmega_q} | \mD, \bOmega_q\right)$ proceed  as described above, with computation of conditional expected outcomes and integration over $\bX|S=P$ using only data from the primary source and the posterior for $\btheta_P^{\bOmega_q}$, the parameters for the primary source under the exchangeability assumption for $\bOmega_q$. 
  
  \subsection{Marginal Likelihoods}
  
  Because the likelihood $p\left(\mD | \bTheta_q, \bOmega_q\right)$ is a product of conditional densities of $Y$ rather than a joint distribution for the observed data, derivation of the marginal likelihood in Equation (\ref{eqn:marginal-likelihood}) treats  $A, S$, and $\bX$ as fixed. For simplicity we derive the marginal likelihood for the primary data source only for the MEM with $Z_h = 0$ for all $h$, indicating no borrowing between data sources. For MEMs with borrowing, data sources with shared parameters $\btheta_P$ are combined for computing (\ref{eqn:marginal-likelihood}).
  
  \subsubsection{Bayesian Linear Model} 
  
   For Model (\ref{eqn:bayes_lm}) with the normal-inverse gamma prior, the marginal likelihood has the multivariate $t$-density \\ $p\left(\bY | \bD\right) \sim t_{2a_P}\left\{\bD\bmu_P, \frac{b_P}{a_P}\left(\bI + \bD\bV_P\bD^T\right)\right\}$, where $\bI$ is the $n_P \times n_P$ identity matrix.

  \subsubsection{BART}
  
  {\it Marginal Likelihood Under Default Prior.} To simplify the derivation, let $\mT_P = \{T_{P1}, \dots, T_{Pm}\}$ and $\mM_P = \{M_{P1}, \dots, M_{Pm}\}$ denote the set of trees and terminal nodes, respectively. The marginal likelihood given model (\ref{eqn:BART-model}) under the default prior specification is
\begin{align*}
    p\left(\bY|\bD\right) = & \sum \int \int p\left(\bY | \bD, \mM_P, \mT_P, \sigsq_P\right) p\left(\mM_P| \mT_P\right)p\left(\mT_P\right)p\left(\sigsq_P\right) d\mM_P d\sigsq_P,
\end{align*}
where the summation is over the discrete distribution of trees. To integrate out $\mM_P$, note that the terminal nodes are {\it a priori} independent and identically distributed as $N\left(0, \tausq\right)$. This implies the prior
$E\left(\bY|\bD, \mT_P\right) \sim N\left(\bs{0}, m\tausq \bR_P\right)$, 
where $\bR_P$ is a correlation matrix with off-diagonal elements $r_{kl}$ equal to the proportion of terminal nodes shared between $E\left(Y_k|A_k, \bX_k\right)$ and $E\left(Y_l|A_l, \bX_l\right)$.
Hence, 
$\bY | \bD, \mT_P, \sigsq_P \sim N\left(\bs{0}, \bSigma_P\right)$, where $\bSigma_P = m\tausq\bR_P + \sigsq_P\bI$. Using this result, we can re-write the marginal likelihood as
\begin{equation*}
 p\left(\bY|\bD\right) \propto \sum \int
 \lvert\bSigma_P\rvert ^ {-\frac{1}{2}}
 e ^ {-\frac{1}{2}\bY^T\bSigma_P ^ {-1} \bY} \left(\frac{1}{\sigsq_P}\right) ^ {\frac{\nu}{2} + 1}
 e ^ {-\frac{\nu\lambda}{2\sigsq_P}} p\left(\mT_P\right) d\sigsq_P,
\end{equation*}
where we have substituted the kernels of the normal probability density function (pdf) for $p\left(\bY|\bD, \mT_P,\sigsq_P\right)$ and the inverse gamma pdf for $p\left(\sigsq_P\right)$. 
At this point we would integrate over $\sigsq_P$, but this is not (analytically) straightforward: $\bSigma_P=m\tausq\bR_P + \sigsq_P\bI$ is a function of $\sigsq_P$, but $\sigsq_P$ cannot be factored out of $\bSigma_P$. For this reason, we revert to the prior on the terminal node values $\mu_{jk} \overset{iid}{\sim} N\left(0, \frac{\sigma^2_P}{\gamma}\right)$ originally used by \citet{Chipman1998} in their Bayesian implementation of classification and regression trees  (see also \citet{Hernandez2018}).


  {\it Marginal Likelihood Under Modified Prior.}
 Assume that the terminal nodes values are {\it a priori} independent and identically distributed as $N\left(0, \frac{\sigsq_P}{\gamma}\right)$, where $\gamma$ is a constant. 
Under this modified prior, the marginal likelihood is
\begin{equation*}
 p\left(\bY|\bD\right) \propto \sum \int
  \lvert\bSigma_P^*\rvert ^ {-\frac{1}{2}}
 e ^ {-\frac{1}{2}\bY^T\bSigma_P^ {*-1} \bY} \left(\frac{1}{\sigsq_P}\right) ^ {\frac{\nu}{2} + 1}
 e ^ {-\frac{\nu\lambda}{2\sigsq_P}} p\left(\mT_P\right) d\sigsq_P,
\end{equation*}
where $\bSigma_P^* = \frac{m\sigsq_P}{\gamma}\bR_P + \sigsq_P\bI$. Note that $\sigsq_P$ can be factored out of $\bSigma_P^*$: $\bSigma_P^* = \sigsq\bU_P$, where $\bU_P = \frac{m}{\gamma}\bR_P + \bI$. The marginal likelihood can now be simplified to
\begin{align*}
    p\left(\bY|\bD\right) 
    \propto &
    \sum \int
    \lvert\sigsq_P \bU_P\rvert ^ {-\frac{1}{2}}
    e ^ {-\frac{1}{2\sigsq_P}\bY^T\bU_P^{-1}\bY}
    \left(\frac{1}{\sigsq_P}\right) ^ {\frac{\nu}{2} + 1}
    e ^ {-\frac{\nu\lambda}{2 \sigsq_P}} 
    p\left(\mT_P\right) d\sigsq_P \\
    = & \sum 
    \lvert\bU_P\rvert ^ {-\frac{1}{2}}
    \left\{
    \int 
    \left(\frac{1}{\sigsq_P}\right) ^ {\frac{n_P + \nu}{2} + 1}
    e ^ {-\frac{1}{2\sigsq_P}\left(\bY ^ T \bU_P ^ {-1} \bY + \nu\lambda\right)}
    d\sigsq_P
    \right\}
    p\left(\mT_P\right) \\
    \propto &
    \sum
    \left[
    \lvert\bU_P\rvert ^ {-\frac{1}{2}}
    \left\{
    1 + \frac{1}{\nu} \bY^T \left(\lambda\bU_P\right) ^ {-1} \bY
    \right\} ^ {-\frac{\nu + n_P}{2}}
    \right]
    p\left(\mT_P\right)
\end{align*}
The term in the square brackets is the density of the central multivariate $t$ distribution with $\nu$ degrees of freedom and shape matrix $\lambda\bU_P$. Note that in the first equality, $\det{\left(\bU_P\right)} ^ {-\frac{1}{2}}$ cannot be treated as a proportionality constant, because $\bR_P$ depends on the structure of the $m$ trees. Simplifying the above expression,
\begin{equation*}
    p\left(\bY|\bD\right) =
    \sum t_{\nu}\left(\bY | \bs{0}, \lambda\bU_P\right)p\left(\mT_P\right).
\end{equation*}
Under the modified prior, the marginal density of $\bY$ is a weighted average of multivariate $t$ densities.

  Although this modified prior is not required for estimation or for computing the marginal likelihood, it does make computing the marginal likelihood more straightforward. Furthermore, BART is traditionally viewed as a ``black box'' method in that in that it takes input and produces output with little intuition for the intermediate steps (but see \citet{Tan2019} for a lucid description of BART). At minimum, this result provides some intuition for how the posterior weights come about, particularly because the $t$-distribution is well-characterized and lends itself to analysis.
\section{Simulation}
 \label{simulation}

  We conducted a simulation study to assess the performance of the BLM and BART while borrowing from a supplemental source. Part 1 demonstrates the borrowing and performance of the estimators under several simple scenarios. Part 2 is designed to test the performance of the estimators under a variety of realistic scenarios. All models were fit using the \verb+borrowr R+ package \citep{borrowr}, which includes an implementation of the BLM and BART using a modified version of the \verb+R BART+ package to draw from the posterior distributions implied by the modified prior on the terminal nodes discussed in Section \ref{sec:BART} \citep{Hernandez2018}.
  
  \subsection{Simulation Part 1}
 
  We consider data from a primary source and a single supplemental source, each with sample size 100, under three scenarios. For all scenarios, $\Pr\left(A = 1\right) = 0.5$ (marginally). For the primary data source the outcome is generated according to the model $Y = A + f\left(X\right) + \epsilon, \epsilon \sim N(0, 1)$, with $f$ and $X$ defined below for each scenario. In all scenarios the PATE is fixed at 1. In the supplemental data source the outcome is generated as $Y = \left(1 + \delta \right)A + f\left(X\right) + \epsilon, \epsilon \sim N(0, 1)$, with $\delta$ varied along a grid of values from -2.5 to 2.5 in increments of 1/2. In all scenarios, the BLM regresses $Y$ on $A$ and $X$ with no interactions or non-linear terms. 
  For BART, we used the default priors $m = 200, k = 2, \nu = 3$, with 100 burn-in posterior draws, and 100 draws from the posterior for inference. The marginal likelihood for BART was estimated by averaging the likelihood over 100 draws from the prior distribution.  
  For each iteration $\gamma$ was set to $\frac{1}{16m\hat{\sigma} ^ 2}$, where $\hat{\sigma}^2$ is estimated from a least squares regression of $Y$ on $A$ and $X$, so that $\gamma\sigma ^ 2$ is comparable to the unmodified BART prior variance of $\frac{1}{16m}$. 
  We include a ``no borrowing'' (NB) estimator that adjusts for confounding but does not borrow from supplemental sources.
  This is a BLM that regresses $Y$ on $A$ and $X$ but uses only the primary data source. 
  We investigated two priors on models. With only one supplemental source, the prior is defined by the prior probability of exchangeability, $\Pr(Z_1 = 1)$. The first prior is $\Pr(Z_1 = 1) = \frac{1}{2}$, and the second is $\Pr(Z_1 = 1) = \left(\frac{1}{2}\right) ^ r$, where $r$ is the number of predictors in the model ($+1$ for the intercept for the BLM). The former is shown in figures with no subscript (``BLM" and ``BART"). The latter is meant to discourage borrowing as the dimension of $X$ grows and is denoted in figures with the subscript $r$ (``BLM$_r$" ``and BART$_r$"). For each scenario we report the estimated bias, root mean squared error (MSE), and the posterior weight corresponding to the MEM in favor of borrowing (that is, parameters between the two data sources are shared) for 1,000 Monte Carlo simulations for each value of $\delta$ within each scenario. 
  
  \subsubsection{Scenario 1: Parallel Linear Response Surfaces}
\begin{figure}
    \centering
    Part 1, Scenario 1
    \includegraphics[width = 1\textwidth]{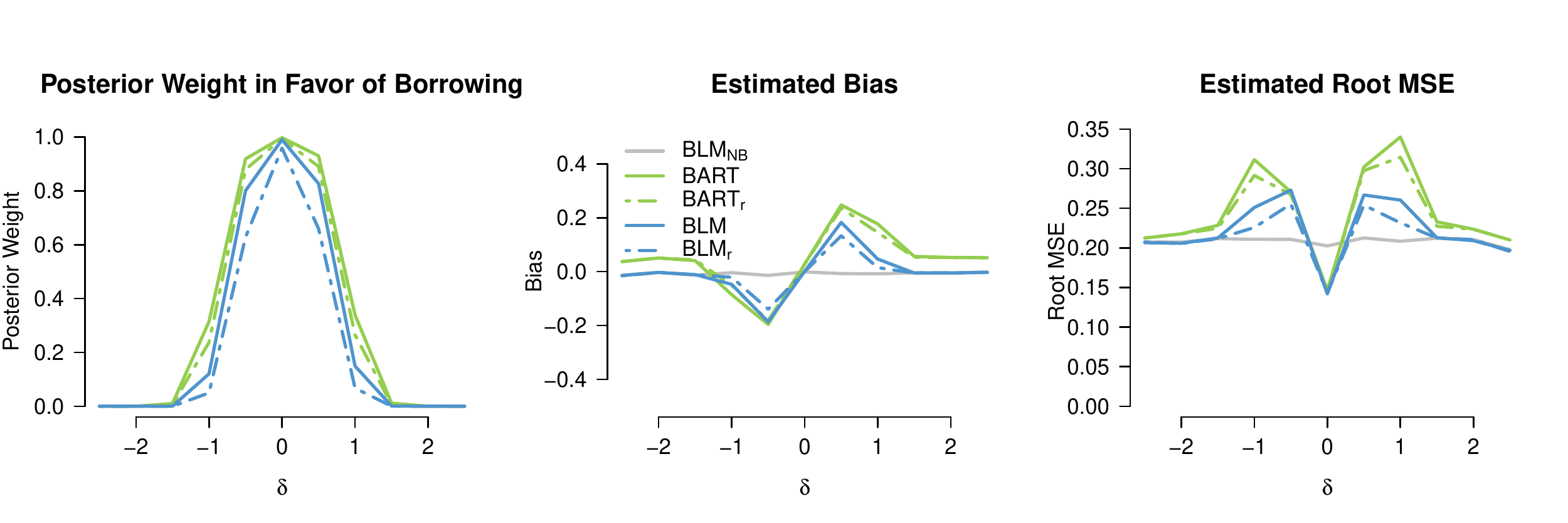}
    Part 1, Scenario 2
    \includegraphics[width = 1\textwidth]{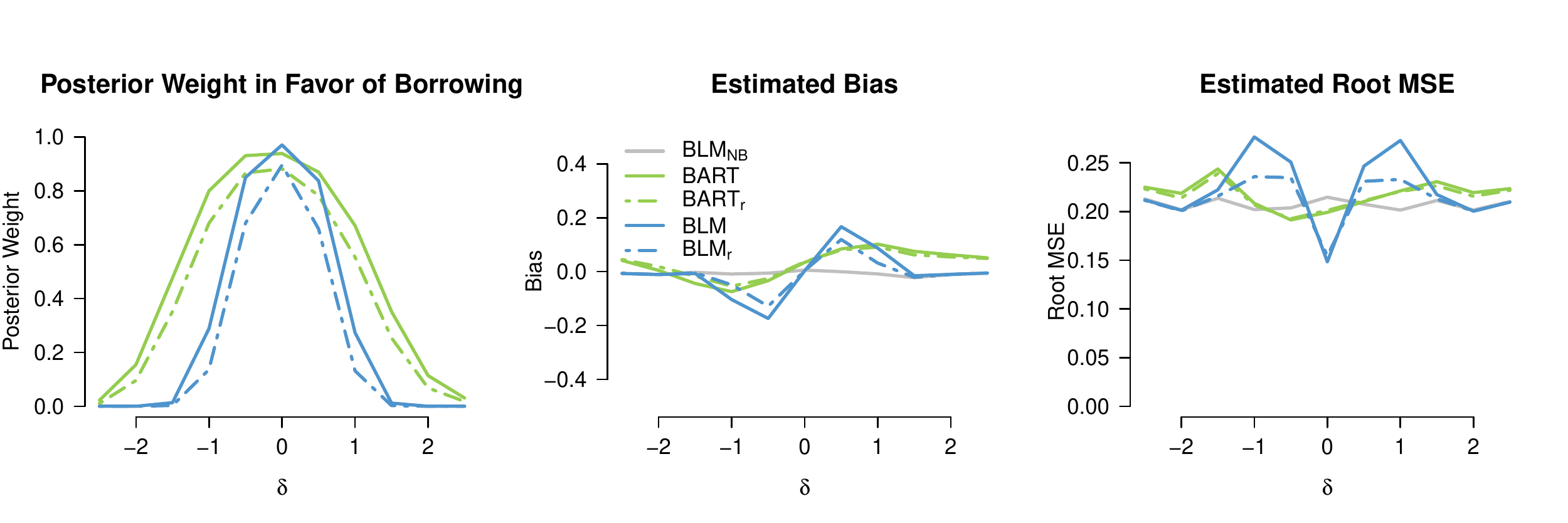}
    Part 1, Scenario 3
    \includegraphics[width = 1\textwidth]{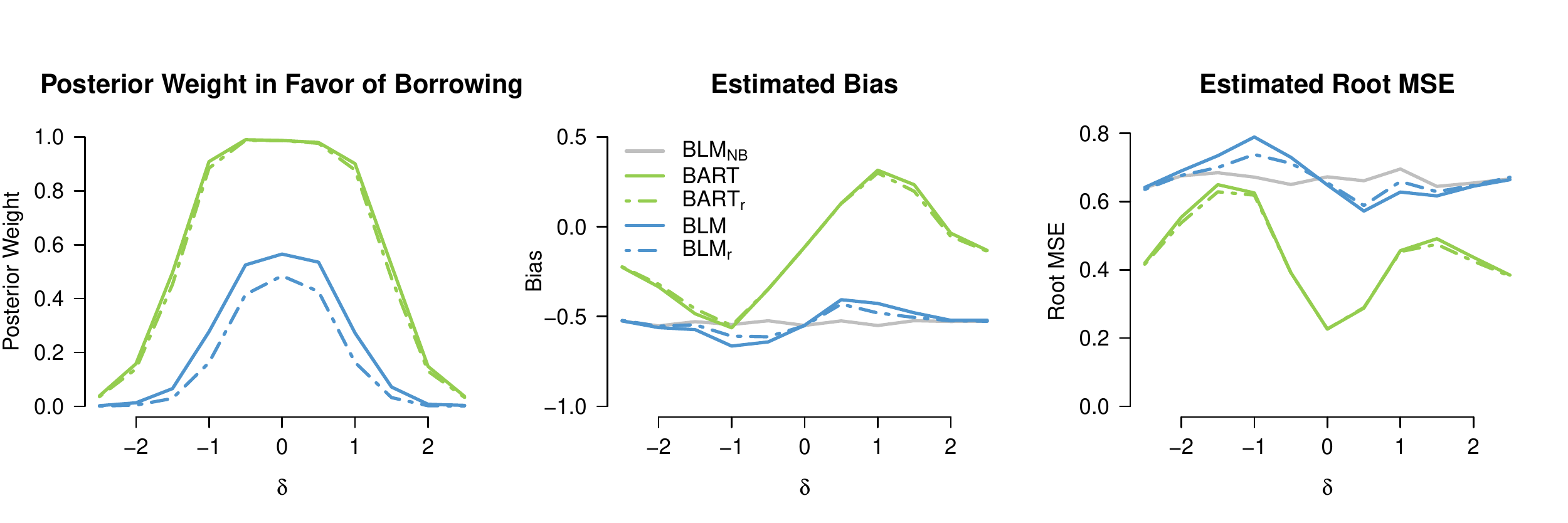}
    \caption{Simulation Part 1 Results. Scenario 1: parallel, linear response surfaces. Scenario 2: different mean of confounder in supplemental source. Scenario 3: over-specified outcome model. $\delta$: Difference in treatment effect between sources (supplemental - primary) on effect size scale. BLM: Bayesian linear model; BLM$_P$: Bayesian linear model with power prior; NB: no borrowing.}
    \label{fig:part-1}
\end{figure}

  For this scenario we generate $X \sim N(0, 1)$, $A|X \sim Ber\left(\left(1 + e ^ {-x}\right) ^ {-1}\right)$, and use $f(x) = x$. The top row of Figure \ref{fig:part-1} shows the results. All estimators except NB borrowing across $\delta$ values of roughly -1.5 to 1.5, with BART borrowing the most and BLM$_P$ the least. All estimators are minimally biased when not borrowing, and show small amounts of bias when borrowing with $\delta \ne 0$. All estimators show roughly the same reduction in MSE when borrowing close to $\delta = 0$. As $\delta$ moves from 0, all estimators begin to show increased MSE relative to MSE when not borrowing, but this penalty is noticeably smaller for BLM$_r$ than for BLM, and the same patter holds for BART$_r$ compared to BART. In general the root MSE for BART is greater than for the BLM.

  \subsubsection{Scenario 2: Parallel Linear Response Surfaces with Different Confounder Distribution in the Supplemental Source}
  
  We generate data for the primary source as for Scenario 1, except that in the supplemental source we generate $X \sim N\left(3, 1\right)$. The purpose of the scenario is to investigate the extent to which an investigator can benefit from including supplemental sources that are drawn from sub-populations with differing distributions of confounders than the primary data source. The middle row of Figure \ref{fig:part-1} shows the results. The performace of BLM is similar to Scenario 1, but BART shows almost no MSE reduction for $\delta$ close to 0, and almost no MSE increase for $\delta$ values farther from 0.

  \subsubsection{Scenario 3: Non-Linear Response Surfaces with Different Conditional Variances in Confounder}
  
  This scenario was informed by previous work \citep{Rubin1973} indicating that linear estimators can perform very poorly when $f(x)$ is non-linear and the conditional variances of $X|A = 0$ and $X|A = 1$ are different. For both data sources, we generate $A \sim Ber(0.5)$ and $X$  according to the conditional distributions $X | A = 0 \sim N\left(0, \frac{4}{3}\right)$ and $X | A = 1 \sim N\left(0, \frac{2}{3}\right)$, where the normal distributions are parameterized with the variance. The outcome is generated using $f(x) = e ^ x$. As shown in in the bottom row of Figure \ref{fig:part-1}, this is a challenging scenario for both estimators. The posterior weights indicate that BART borrows across a wide range of $\delta$, while the weights for the BLM are lower at $\delta = 0$ than in both previous scenarios. The BLM is biased and the MSE is higher than BART for all values of $\delta$.  Although negative bias is evident for BART at $\delta = 0$, the smaller magnitude bias and MSE of BART clearly indicates that BART performs better than the BLM in this scenario.

  \subsection{Simulation Part 2: Atlantic Causal Inference Competition Simulation}
  
  In the second part of the simulation, we applied the estimators to data from the 2016 Atlantic Causal Inference Conference (ACIC) as described by \citet{Dorie2019}. The goal of the part of the simulation is to assess the performance of the estimators under a variety of data-generating scenarios more realistic than those presented in part 1. Briefly, ACIC generates simulated data according to 77 scenarios, with simulation parameters varied across 6 ``knobs": degree of non-linearity in the response surface; degree of non-linearity in the treatment surface; proportion treated; overlap, defined as the degree to which areas of the covariate space have only untreated individuals; alignment, defined as the degree to which covariates appear in both the treatment and the response model; and treatment effect heterogeneity, or departures from parallel response surfaces between treated and untreated observations. Covariates are from a real data set and include 3 categorical, 5 binary, 27 count, and 23 continuous variables. See \citet{Dorie2019} for additional details. The design matrix with dummy coding for categorical predictors had 81 predictors excluding the intercept term. For each iteration we randomly selected 400 observations and assigned 200 to be primary and 200 to be supplemental sources. The first 100 in each were treated, and the remaining were control. 
  
  In addition to the model priors described above, we included for BART only additional model priors that discouraged borrowing to varying degrees as a function of $r$, the number of predictors: BART$_{log2r}$, with model prior $\Pr(Z_1 = 1) = \left(\frac{1}{2}\right) ^ {\log_2 r} = \frac{1}{r}$; and BART$_{r2}$, with model prior $\Pr(Z_1 = 1) = \left(\frac{1}{2}\right) ^ {\frac{r}{2}}$. Pilot work suggested that BART$_r$, with model prior $\Pr(Z_1 = 1) = \left(\frac{1}{2}\right) ^ r$ discouraged borrowing too strongly in this high-dimensional setting, so we excluded BART with this model prior. Futhermore, the performance of this implmentaion of the BLM in this simulation was uniformly inferior to BART, so we show results considered BLM$_r$ with no other model priors. We again included a ``no borrowing'' estimator that uses only the primary day. For this part the no borrowing estimator is denoted BART$_{NB}$.  All potential predictors were included for BART. The BLM regressed the outcome on all variables without considering possible interactions, non-linear, or polynomial terms.  We increased the number of posterior draws to 1,000 due to the more complicated response surfaces compared to part 1. 

\begin{figure}
    \centering
    \includegraphics[width = 1\textwidth]{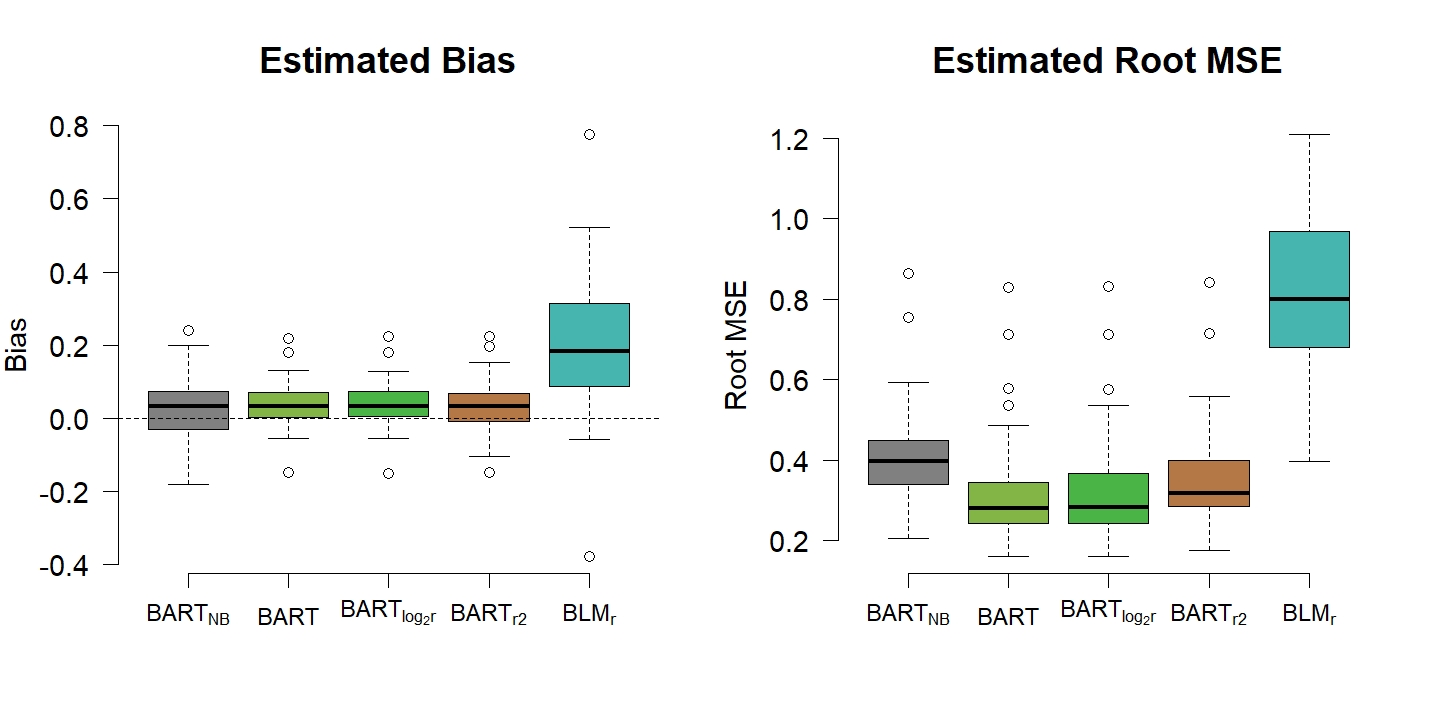}
    Part 1, Scenario 2
    \caption{ACIC Simulation Results. Each point is the mean for 50 Monte Carlo iterations for one data-generating scenario. NB: no borrowing. Other subscripts indicate different model prior probabilities.}
    \label{fig:acic-zero-delta}
\end{figure}

  Figure \ref{fig:acic-zero-delta} shows the results, with each point in the boxplots showing the mean over 50 Monte Carlo iterations. BART is minimally biased when not borrowing and when borrowing under each of the model priors, but the BLM is signifcantly positively biased. All BART estimator that borrow show reduced root MSE compared to BART$_{NB}$, with smaller reductions seen for model priors that more strongly discourage borrowing. In contrast, the root MSE of BLM$_r$ is quite high and performs much worse than all BART estimators.
  
\begin{figure}
    \centering
    \includegraphics[width = 1\textwidth]{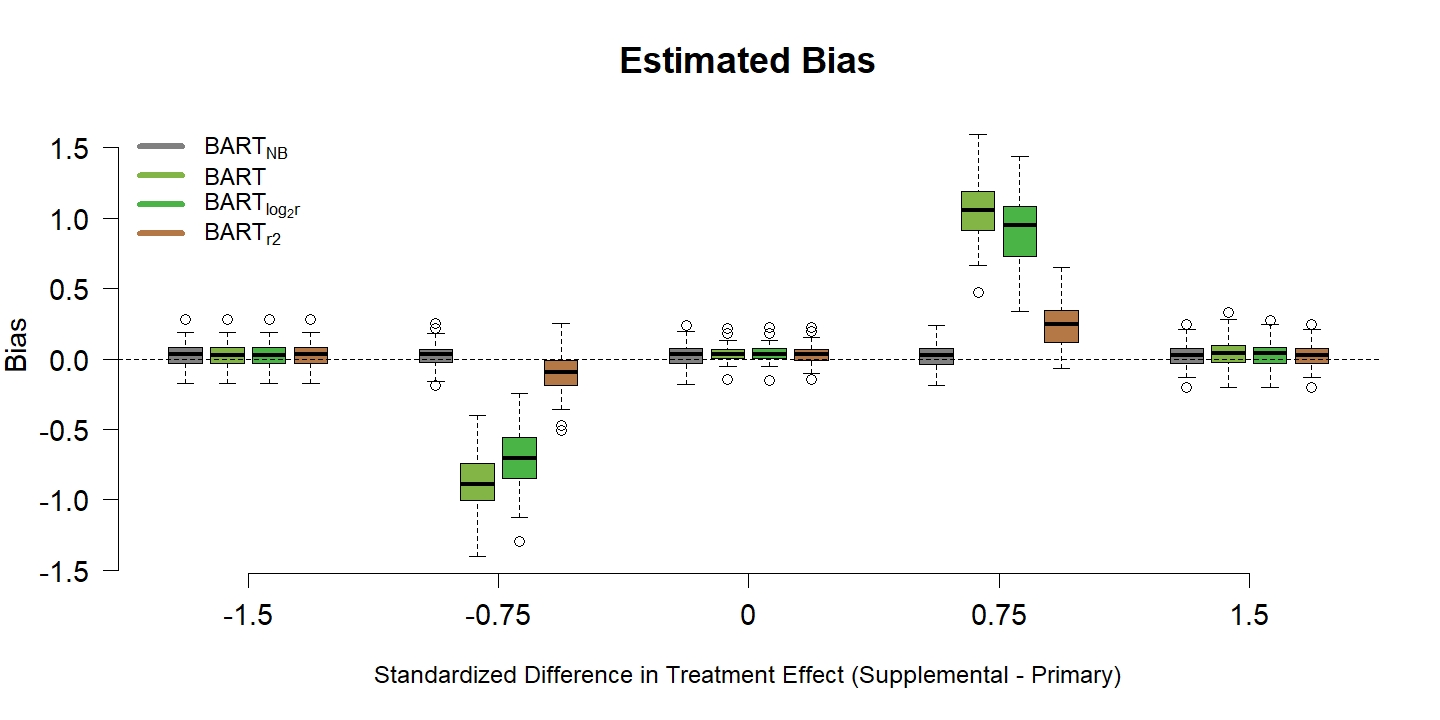}
    \includegraphics[width = 1\textwidth]{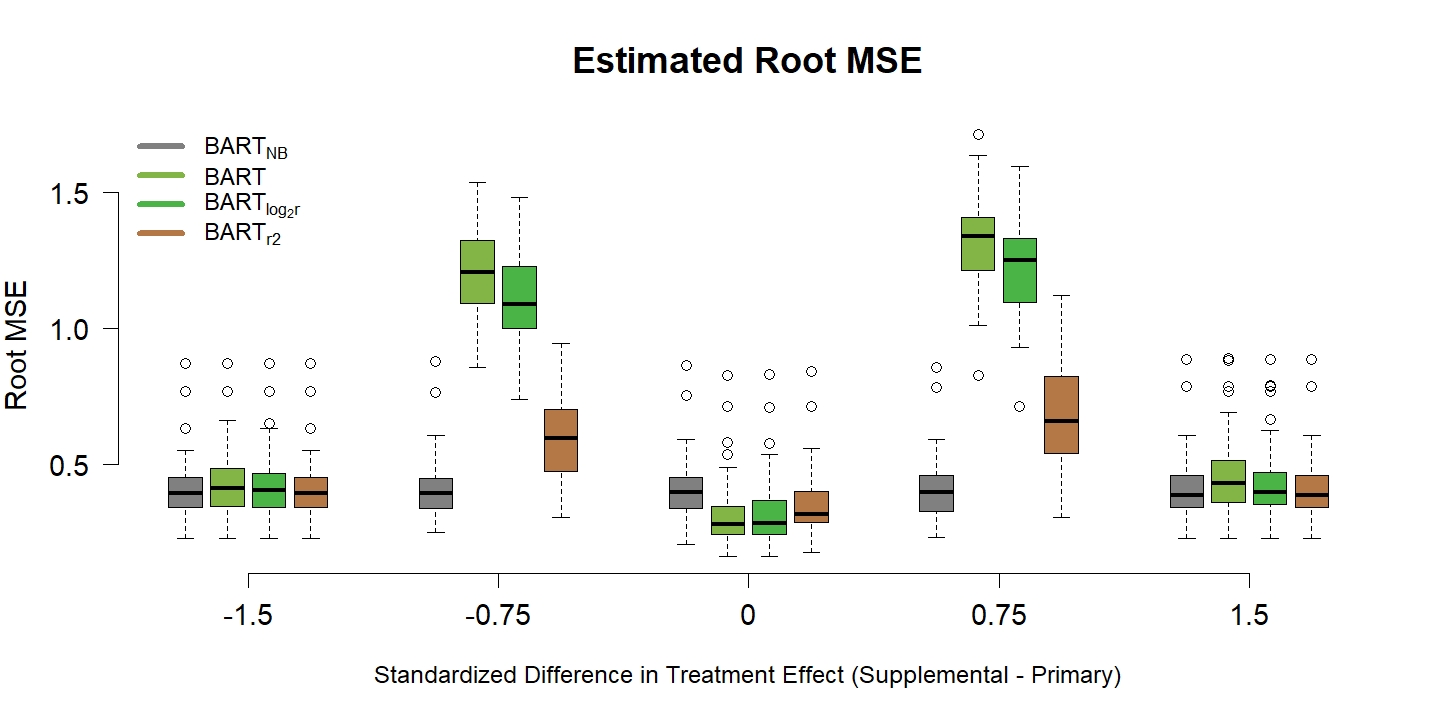}
    \caption{ACIC Simulation Results. Each point is the mean for 50 Monte Carlo iterations for one data-generating scenario. NB: no borrowing. Other subscripts indicate different model prior probabilities.}
    \label{fig:acic-all-delta}
\end{figure}

  We also varied the treatment effect in the supplemental source to to evaluate the performance of the model priors. Similar to part 1, we varied the treatment effect in the supplemental source by adding $\delta \times sd(Y)$, where $sd(Y)$ is the standard deviation of the outcome before modifying the treatment effect. $\delta$ was varied along a grid from -1.5 to 1.5 increments of 0.75. Because the BLM performed poorly in Figure \ref{fig:acic-zero-delta}, we consider only BART with various model priors. Figure \ref{fig:acic-all-delta} shows the results. Note that the boxplots at 0 simply duplicate the results in Figure \ref{fig:acic-zero-delta}. When the standardized difference in treatment effect is $\pm 1.5$, there is no borrowing and the performance of all estimators is roughly equivalent. At $\pm 0.75$, increased bias and higher root MSE is evident. The magnitude of bias and root MSE are higher using model priors that do not discourage borrowing based on the the number of predictors (BART), and lowest for models that do (BART$_{r2}$), with BART$_{log2r}$ showing intermediate magnitude of bias and root MSE.  
  
  \subsection{Simulation Summary}
  
  The simulation results suggest a simple conclusion. Part 1 suggests that if we are reasonably confident we can specify a correct, or approximately correct, outcome model, BLM$_r$ is the best estimator. Part 2 suggests that if the dimension of the predictors is large, or if we're not confident we can correctly specify the outcome model, BART is the best estimator. Furthermore, part 2 suggests better performance using model priors that discourage borrowing as the dimension of the predictors grows. Although the potential root MSE reduction is limited with these priors, the potential root MSE penalty is much lower as well.

\section{Application} \label{application}

We analyzed data from 3 RCTs evaluating the effect of VLNC cigarettes on cigarettes smoked per day (CPD), using CENIC project 1, study 2 (P1S2) as the primary source and project 1, study 1 (P1S1) and project 2 (P2) as supplemental sources. P1S1 was a 6-week trial that randomly assigned participants to one of seven groups, consisting of a usual brand group and groups assigned experimental cigarettes with nicotine content ranging from 15.8 mg nicotine/gram tobacco to 0.4 mg nicotine/gram tobacco, or a group assigned high-tar cigarettes with 0.4 mg nicotine/gram tobacco (hereafter abbreviated mg/g)  \citep{Donny2015}. For this application the 15.8 mg/g group is considered control and the 0.4 mg/g groups, high and low tar combined, are considered treatment. P1S2 was a 6-week, 2x2 factorial RCT that randomly assigned current smokers to smoke experimental cigarettes with nicotine content of 15.8 mg/g or 0.4 mg/g, with or without nicotine replacement therapy (NRT) \citep{Smith2019}. For this application, the 15.8 mg/g  group without NRT is the control group, and the 0.4 mg/g group without NRT is the treatment group. In contrast to P1S1 and P1S2, P2 was a 20-week RCT designed to test gradual vs. immediate nicotine reduction \citep{Hatsukami2018}.  Control participants were assigned to smoke experimental cigarettes with nicotine content of 15.5 mg/g; gradual reduction participants were assigned to smoke experimental cigarettes with nicotine content gradually reduced from 15.5 mg/g to 0.4 mg/g over the trial, and the immediate reduction group was assigned to immediately begin smoking cigarettes with nicotine content of 0.4 mg/g. Participants were assessed mid-trial at 8 weeks rather than 6 as in P1S1 and P1S2, so we use the outcome data from week 8. For this application the 15.5 mg/g group was treated as control, and the immediate reduction group was treated as the treatment group. 

In these 3 trials, treatment was randomly assigned. However, causal effect estimators that adjust for confounding are still necessary due to noncompliance: participants self-reported whether they smoked non-study cigarettes, and if so, were considered noncompliant. Although treatment assignment is randomized, compliance to treatment is not, and treatment-control comparisons among compliant participants are therefore confounded. We considered participants compliant if they reported smoking 0 non-study cigarettes per day and if their total nicotine equivalents was less than 6.41, as values exceeding this threshold are inconsistent with compliance \citep{denliner2016}. Participants in control groups are smoking investigational cigarettes, but these contain normal amounts of nicotine. Participants in control groups who smoke non-study cigarettes are noncompliant from a strict perspective, but as the study cigarettes are similar to usual brand cigarettes, we consider all participants in the control groups compliant and take their outcome to be study CPD plus non-study CPD.

The estimators described can accommodate this situation with minimal changes. Let $Q$ be the number of non-study cigarettes smoked per day. A participant if considered compliant if $Q = 0$ and non-compliant otherwise. Consider expanded notation for the potential outcomes, with $\Ystar(a, s, q)$ indicating the potential outcome if, possibly contrary to fact, $A = a, S = s$, and $Q = q$. Also define the compliance indicator $C = I\left(Q = 0\right)$. Our target of inference is the expected difference in the number of study cigarettes smoked per day for treatment versus control in the primary data source assuming all participants were to be compliant, or $ \Delta_P = E_{\bX|S = P}\left[E \left\{\Ystar(1, P, 0) - \Ystar(0, P, 0) |\bX, S = P\right\}\right]$. Valid inference requires the following modifications to the assumptions in Section \ref{sec:assumptions}. We assume that if $A = a, S = s$ and $Q = 0$, then $Y = \Ystar(a, s, 0)$, that is, if a participant is compliant with the study protocol, then her observed outcome is identical to the potential outcome for $q = 0$ (consistency assumption).  We further assume that we have collected sufficient detail on participant characteristics so that compliance is conditionally ignorable, or $\left.\left\{C \perp \Ystar(0,S,0), \Ystar(1,S,0)\right\} \;\middle\vert\; \bX, S\right.$ (conditional ignorability and no unmeasured confounders).  Under these assumptions we have $E\left\{\Ystar(1, S, 0) | S, \bX\right\} = E(Y | A = 1, C = 1, S, \bX)$ and $E\left\{\Ystar(0, S, 0) | S, \bX\right\} = E(Y | A = 0, C = 1, S, \bX)$. The expectations on the right side of the equalities can be estimated with the BLM and BART as described above.

To identify variables associated with compliance, we fit a lasso regression model \citep{Tibshirani1996} with treatment indicators for treatment ($A$) and compliance ($C$); dummy variables for data source, age, race, educational attainment; and a variety of other potential confounders $X$ including variables measuring level of addiction, level of dependence, and satisfaction with and craving for usual brand cigarettes. The model indicated non-zero coefficients for $A$, $C$, source, age, baseline CPD, and dependence (Fagerstrom Test for Nicotine Dependence; FTND).
To identify possible interactions, we next fit a least squares model with additive linear terms for these variables as well as all pairwise interactions excluding terms involving a treatment-compliance interaction. (An interaction between $A$ and $C$ cannot be evaluated in this case because all participants in the control group were treated as compliant.) The model indicated significant interactions between $A$ and age, between $C$ and baseline CPD, and between $A$ and baseline CPD. The Bayesian model used additive linear terms for these variables as well as the interactions identified in the last squares model. For BART, we used $A, C$, and all baseline variables with no screening for variable selection. 

Prior to implementing the proposed estimators, we fit the BLM and BART separately for each data source and estimated the causal effect of treatment within each source. This step is not required, but we report these results to provide intuition for whether borrowing from P1S1 and P2 is appropriate. Table \ref{tab:coefficients} summarizes the samples sizes (after removing observations with missing data for outcomes and potential confounders), BLM coefficients, and causal effects estimated separately within each data source. The model coefficients are similar between data sources, indicating that borrowing from both supplementary sources may be appropriate. In addition, the causal effects estimated within each source are also similar. Although the BLM and BART borrow based on assuming conditional exchangeability, and not on the marginal causal effects, the similarity in the causal effects nevertheless makes us more confident that borrowing from P1S1 and P2 is appropriate.  


\begin{table}[!ht]
    \centering
    \caption{Sample Sizes, Posterior mean (sd) of the Bayesian linear model coefficients, and estimated treatment effects ($\Delta$) estimated for each data source without borrowing.}
    \begin{tabular}{lccc}
    \hline
& {\bf P1S2} & {\bf P1S1} & {\bf P2} \\
\hline
{\bf Sample Sizes} \\
\quad Control & 54 & 105 & 216\\
\quad Treatment & 48 & 218 & 375\\
\quad Total & 102 & 323 & 591\\
{\bf Coefficients} \\
\quad Intercept & 18.22(0.65) & 16.27(0.44) & 16.52(0.29)\\
\quad $A$ & -16.95(7.18) & -13.39(4.07) & -11.14(2.56)\\
\quad $C$ & -8.41(4.94) & -7.20(3.04) & -1.89(1.69)\\
\quad Age & 0.00(0.08) & -0.01(0.06) & 0.01(0.04)\\
\quad Baseline CPD & 0.52(0.27) & 0.33(0.22) & 0.80(0.11)\\
\quad FTND & 0.63(0.41) & 0.49(0.24) & 0.51(0.17)\\
\quad $A \times$ Age & 0.20(0.12) & 0.07(0.07) & 0.16(0.05)\\
\quad $C \times$ Baseline CPD & 0.58(0.24) & 0.64(0.20) & 0.25(0.09)\\
\quad $A \times$ Baseline CPD & 0.06(0.22) & 0.33(0.21) & -0.09(0.08)\\
$\bf{\bs{\Delta_P}}$ {\bf (Treatment - Control)} & \\
\quad BLM & -6.35(2.19) & -5.37(1.43) & -5.03(0.75)\\
\quad BART & -5.91(2.02) & -6.04(1.29) & -4.96(0.75)\\

\hline
\end{tabular}
    \label{tab:coefficients}
\end{table}

For models with borrowing we assumed the model prior $\Pr(Z_h = 1) = \left(\frac{1}{2}\right) ^ r$ with $r = 39$ predictors. Table \ref{tab:estimates} shows the results of the primary analysis. For the BLM, the the most weight is given to models assuming both P1S1 and P2 are exchangeable with P1S2, and assuming only P2 is. In contrast, BART essentially indicates that only P2 is exchangeable with P12. From the BLM, the posterior mean(SD) reduction in study cigarettes smoked per day for being assigned to treatment is 5.08(0.68) as estimated from the BLM; the BART estimate is very similar at 4.91(0.70), suggesting that the BLM is sufficiently complex to model the response surface. Both estimators show dramatic reductions in the posterior variance of the causal effect compared to no borrowing. It is counterintuitive that the BART estimated causal effect is not within the range of effects separated separately within each source, but as borrowing is on conditional means and not on marginal means, it will not necessarily be the case that the marginal effect will be shrunk toward either marginal treatment effect.


\begin{table}[!ht]
    \centering
    \caption{Posterior model weights ($\omega$) and mean(sd) of the treatment effect ($\Delta_P$) for the Bayesian linear model (BLM) and Bayesian additive regression trees (BART).}
    \label{tab:estimates}
    \begin{tabular}{cccccc}
\hline
\multicolumn{6}{c}{\textbf{\emph{Exchangeability and Posterior MEM Weights}}} \\
\hline
& \multicolumn{2}{c}{Exchangeable with P1S2} && \multicolumn{2}{c}{$\omega$}\\
\cmidrule{2-3} \cmidrule{5-6}
MEM & P1S1 & P2 && BLM & BART\\
 \hline
1 & Yes & Yes && 0.6343 & 0.0003\\
2 & No & Yes  && 0.3652 & 0.9996\\
3 & Yes & No  && 0.0005 & 0.0000\\
4 & No & No   && 0.0000 & 0.0001\\
 \hline
 \multicolumn{6}{c}{\textbf{\emph{Weighted Posterior Mean(SD) Treatment Effects}}}\\
 \hline
 &&&& BLM & BART \\
 \cmidrule{5-6}
 \multicolumn{3}{c}{$\Delta_P$ (Treatment - Control)} && -5.08(0.68) & -4.91(0.70)\\
 \hline
 \end{tabular}
\end{table}
\section{Discussion} \label{discussion}

  In this manuscript we demonstrated how to borrow from supplementary data sources to estimate causal effects from a primary data source. The simulation demonstrates that both the BLM and BART are viable estimators with good properties depending on the level of confounding and on the complexity of the response surface. The primary advantages of the BLM are that it results in interpretable coefficients, and that the variance of the causal estimator may be lower than for BART due to its frequentist operating characeristics. The primary advantage of BART is that the analyst is not required to specify a function for for the covariate-response association, meaning that BART can naturally capture complicated relationships among the covariates and outcome. 

  The primary drawback of our proposed estimators is that they rely on regression models to estimate causal effects. A major disadvantage of this is that, in cases with incomplete overlap in the covariate distribution between treatment and control groups, the estimated response surfaces in many areas are based on extrapolation \citep{lunceford2004}. For this reason, regression models are sometimes avoided for causal inference, and other methods, particularly those based on propensity scores, are favored. 
  One of the primary advantages of the propensity score is that it allows the analyst to condition on the propensity score in the outcome model rather than specifying a complete covariate-outcome model. This is problematic when attempting to borrow from supplemental sources, however. Borrowing between data sources happens most naturally in a Bayesian setting, but propensity score use in the Bayesian paradigm is not straightforward.
  Conditioning on a propensity score in the outcome model in a Bayesian design does not correspond to a valid use of Bayes theorem with correct posterior inference \citep{Zigler2016}.   
  
  We chose to borrow based on the similarity between regression coefficients and parameters, and not on the causal effect itself. If we had access to the individual-level causal effects, we could avoid the use of regression models entirely and base our inference and borrowing on the distribution of difference scores. Unfortunately, such data are not available as we can only observe one potential outcome, and never both. Furthermore, individual-level unobserved potential outcomes cannot be predicted without making assumptions about unidentifiable parameters. Nevertheless, it may be possible to use a model to estimate mean causal effects and base the borrowing on the mean potential outcome difference scores. We leave this possibility open for future consideration. We also developed model priors that discourage borrowing as the number of predictors grows, but optimizing this prior to minimize potential MSE increases requires additional investigation.
  
  In summary, we proposed showed how to use two existing modelling strategies and showed how to borrow from supplemental data sources using MEMs. The estimators can even be used to borrow from observation data to estimate causal effects in an RCT provided that the share predictors of the outcome. Both the BLM and BART estimators performed well under simulation scenarios, and the estimators gave sensible results in the application, dramatically reducing the variance of the causal effect estimator. Throughout, we assume that we have a primary data source and supplemental sources. We plan to extend our estimators to cases with symmetrical borrowing, where no source is considered primary, but we wish to borrow between sources as appropriate to estimate causal effects.
  
 \section*{Acknowledgements}

This research was partially funded by NIH under grants P30-CA077598, R01-CA214824, and R01-CA225190 from the National Cancer Institute and R03-DA041870, R01-DA046320 and U54-DA031659 from the National Institute on Drug Abuse and FDA Center for Tobacco Products (CTP). The content is solely the responsibility of the authors and does not necessarily represent the official views of the NIH or FDA CTP.

\section*{Simulation Code and Software}

Simulation code is available at \url{https://github.com/jeffrey-boatman/asymmetric-borrowing-simulation}. The \verb+R+ package \verb+borrowr+ implements the estimators described in this manuscript and is available on the Comprehensive R Archive Network. Research data are not shared.

\bibliographystyle{biom} 
\bibliography{refs}

\begin{thebibliography}{99}

\bibitem[Boatman \emph{and others}(2019)Boatman, Vock and Koopmeiners]{borrowr}
\textsc{Boatman, Jeffrey~A., Vock, David~M. and Koopmeiners, Joseph~S.} (2019).
\newblock {\em borrowr: Estimate Causal Effects with Borrowing Between Data
  Sources\/}.
\newblock R package version 0.1.1.

\bibitem[Chipman \emph{and others}(1998)Chipman, George and
  McCulloch]{Chipman1998}
\textsc{Chipman, Hugh~A., George, Edward~I. and McCulloch, Robert~E.} (1998).
\newblock {Bayesian CART model search}.
\newblock {\em Journal of the American Statistical
  Association\/}~\textbf{94}(443), 935--948.

\bibitem[Chipman \emph{and others}(2010)Chipman, George and
  McCulloch]{Chipman2010}
\textsc{Chipman, Hugh~A., George, Edward~I. and McCulloch, Robert~E.} (2010).
\newblock {BART: Bayesian additive regression trees}.
\newblock {\em Annals of Applied Statistics\/}~\textbf{4}(1), 266--298.

\bibitem[Denlinger \emph{and others}(2016)Denlinger, Smith, Murphy,
  Koopmeiners, Benowitz, Hatsukami, Pacek, Colino, Cwalina and
  Donny]{denliner2016}
\textsc{Denlinger, R.~L., Smith, T.~T., Murphy, S.~E., Koopmeiners, J.~S.,
  Benowitz, N.~L., Hatsukami, D.~K., Pacek, L.~R., Colino, C., Cwalina, S.~N.
  and Donny, E.~C.} (2016).
\newblock Nicotine and anatabine exposure from very low nicotine content
  cigarettes.
\newblock {\em Tobacco Regulatory Science\/}~\textbf{2}(2), 186--203.

\bibitem[Donny \emph{and others}(2015)Donny, Denlinger, Tidey, Koopmeiners,
  Benowitz, Vandrey  et~al.]{Donny2015}
\textsc{Donny, Eric~C., Denlinger, Rachel~L., Tidey, Jennifer~W., Koopmeiners,
  Joseph~S., Benowitz, Neal~L., Vandrey  \emph{and others}}. (2015).
\newblock {Randomized Trial of Reduced-Nicotine Standards for Cigarettes}.
\newblock {\em New England Journal of Medicine\/}~\textbf{373}(14), 1340--1349.

\bibitem[Dorie \emph{and others}(2019)Dorie, Hill, Shalit, Scott and
  Cervone]{Dorie2019}
\textsc{Dorie, Vincent, Hill, Jennifer, Shalit, Uri, Scott, Marc and Cervone,
  Dan}. (2019).
\newblock {Automated versus Do-It-Yourself Methods for Causal Inference:
  Lessons Learned from a Data Analysis Competition}.
\newblock {\em Statist. Sci.\/}~\textbf{34}(1), 43--68.

\bibitem[Gelman(2006)Gelman]{Gelman2006}
\textsc{Gelman, Andrew}. (2006).
\newblock {Prior distributions for variance parameters in hierarchical
  models(Comment on Article by Browne and Draper)}.
\newblock {\em Bayesian Analysis\/}~\textbf{1}(3), 515--534.

\bibitem[Hatsukami \emph{and others}(2018)Hatsukami, Luo, Jensen, Al'Absi,
  Allen, Carmella  et~al.]{Hatsukami2018}
\textsc{Hatsukami, Dorothy~K., Luo, Xianghua, Jensen, Joni~A., Al'Absi,
  Mustafa, Allen, Sharon~S., Carmella, Steven~G.  \emph{and others}}. (2018,
  sep).
\newblock {Effect of immediate vs gradual reduction in nicotine content of
  cigarettes on biomarkers of smoke exposure a randomized clinical trial}.
\newblock {\em JAMA - Journal of the American Medical
  Association\/}~\textbf{320}(9), 880--891.

\bibitem[Hern{\'{a}}ndez \emph{and others}(2018)Hern{\'{a}}ndez, Raftery,
  Pennington and Parnell]{Hernandez2018}
\textsc{Hern{\'{a}}ndez, Belinda, Raftery, Adrian~E., Pennington, Stephen~R.
  and Parnell, Andrew~C.} (2018).
\newblock {Bayesian Additive Regression Trees using Bayesian model averaging}.
\newblock {\em Statistics and Computing\/}~\textbf{28}, 869--890.

\bibitem[Hill(2011)Hill]{Hill2011}
\textsc{Hill, Jennifer~L}. (2011).
\newblock {Bayesian Nonparametric Modeling for Causal Inference}.
\newblock {\em Journal of Computational and Graphical
  Statistics\/}~\textbf{20}(1), 217--240.

\bibitem[Hobbs \emph{and others}(2011)Hobbs, Carlin, Mandrekar and
  Sargent]{Hobbs2011}
\textsc{Hobbs, Brian~P., Carlin, Bradley~P., Mandrekar, Sumithra~J. and
  Sargent, Daniel~J.} (2011).
\newblock {Hierarchical Commensurate and Power Prior Models for Adaptive
  Incorporation of Historical Information in Clinical Trials}.
\newblock {\em Biometrics\/}~\textbf{67}, 1047--1056.

\bibitem[Ibrahim and Chen(2000)Ibrahim and Chen]{Ibrahim2000}
\textsc{Ibrahim, Joseph~G. and Chen, Ming-Hui}. (2000).
\newblock {Power prior distributions for regression models}.
\newblock {\em Statistical Science\/}~\textbf{15}(1), 46--60.

\bibitem[Kaizer \emph{and others}(2018)Kaizer, Koopmeiners and
  Hobbs]{Kaizer2018}
\textsc{Kaizer, Alexander~M., Koopmeiners, Joseph~S. and Hobbs, Brian~P.}
  (2018).
\newblock {Bayesian hierarchical modeling based on multisource
  exchangeability}.
\newblock {\em Biostatistics\/}~\textbf{19}(2), 169--184.

\bibitem[Lunceford and Davidian(2004)Lunceford and Davidian]{lunceford2004}
\textsc{Lunceford, J.~K. and Davidian, M.} (2004).
\newblock Stratification and weighting via the propensity score in estimation
  of causal treatment effects: a comparative study.
\newblock {\em Statistics in Medicine\/}~\textbf{23}, 2937--2960.

\bibitem[Nardone \emph{and others}(2016)Nardone, Donny, Hatsukami, Koopmeiners,
  Murphy, Strasser, Tidey, Vandrey and Benowitz]{Nardone2016}
\textsc{Nardone, Natalie, Donny, Eric~C., Hatsukami, Dorothy~K., Koopmeiners,
  Joseph~S., Murphy, Sharon~E., Strasser, Andrew~A., Tidey, Jennifer~W.,
  Vandrey, Ryan and Benowitz, Neal~L.} (2016).
\newblock {Estimations and predictors of non-compliance in switchers to reduced
  nicotine content cigarettes}.
\newblock {\em Addiction\/}~\textbf{111}(12), 2208--2216.

\bibitem[Nethery \emph{and others}(2019)Nethery, Mealli and
  Dominici]{Nethery2019}
\textsc{Nethery, Rachel~C., Mealli, Fabrizia and Dominici, Francesca}. (2019).
\newblock {Estimating population average causal effects in the presence of
  non-overlap: The effect of natural gas compressor station exposure on cancer
  mortality}.
\newblock {\em Annals of Applied Statistics\/}~\textbf{13}(2), 1242--1267.

\bibitem[Raftery \emph{and others}(1997)Raftery, Madigan and
  Hoeting]{Raftery1997}
\textsc{Raftery, Adrian~E, Madigan, David and Hoeting, Jennifer~A}. (1997).
\newblock {Bayesian Model Averaging for Linear Regression Models}.
\newblock {\em Journal of the American Statistical
  Association\/}~\textbf{92}(437), 179--191.

\bibitem[Robins and Hern\'an(2009)Robins and Hern\'an]{robins2008}
\textsc{Robins, James~M. and Hern\'an, Miguel~A.l}. (2009).
\newblock {\em Longidtudinal Data Analysis\/}, Chapter 23: Estimation of the
  {C}ausal {E}ffects of {T}ime-{V}arying {E}xposures. Boca Raton: Chapman \&
  Hall/CRC.

\bibitem[Rubin(1973)Rubin]{Rubin1973}
\textsc{Rubin, Donald~B.} (1973).
\newblock {The Use of Matched Sampling and Regression Adjustment to Remove Bias
  in Observational Studies}.
\newblock {\em Biometrics\/}~\textbf{29}, 185--203.

\bibitem[Rubin(1981)Rubin]{Rubin1981}
\textsc{Rubin, Donald~B.} (1981).
\newblock {The Bayesian Bootstrap}.
\newblock {\em The Annals of Statistics\/}~\textbf{9}(1), 130--134.

\bibitem[Smith \emph{and others}(2019)Smith, Koopmeiners, Tessier, Davis,
  Conklin, Denlinger-Apte  et~al.]{Smith2019}
\textsc{Smith, Tracy~T., Koopmeiners, Joseph~S., Tessier, Katelyn~M., Davis,
  Esa~M., Conklin, Cynthia~A., Denlinger-Apte, Rachel~L.  \emph{and others}}.
  (2019, oct).
\newblock {Randomized Trial of Low-Nicotine Cigarettes and Transdermal
  Nicotine}.
\newblock {\em American Journal of Preventive Medicine\/}~\textbf{57}(4),
  515--524.

\bibitem[Tan and Roy(2019)Tan and Roy]{Tan2019}
\textsc{Tan, Yaoyuan~Vincent and Roy, Jason}. (2019, aug).
\newblock {Bayesian additive regression trees and the General BART model}.
\newblock {\em Statistics in Medicine\/}~\textbf{38}, 5048--5069.

\bibitem[Tibshirani(1996)Tibshirani]{Tibshirani1996}
\textsc{Tibshirani, Robert}. (1996, jan).
\newblock {Regression Shrinkage and Selection Via the Lasso}.
\newblock {\em Journal of the Royal Statistical Society: Series B
  (Methodological)\/}~\textbf{58}(1), 267--288.

\bibitem[Wang \emph{and others}(2015)Wang, Dominici, Parmigiani and
  Zigler]{Wang2015}
\textsc{Wang, Chi, Dominici, Francesca, Parmigiani, Giovanni and Zigler,
  Corwin~Matthew}. (2015).
\newblock {Accounting for uncertainty in confounder and effect modifier
  selection when estimating average causal effects in generalized linear
  models}.
\newblock {\em Biometrics\/}~\textbf{71}, 654--665.

\bibitem[Zigler(2016)Zigler]{Zigler2016}
\textsc{Zigler, Corwin~Matthew}. (2016).
\newblock {The Central Role of Bayes' Theorem for Joint Estimation of Causal
  Effects and Propensity Scores}.
\newblock {\em The American Statistician\/}~\textbf{70}(1), 47--54.

\end{thebibliography}

\end{document}